\documentclass[aps,pra,showpacs,twocolumn]{revtex4-1}
\usepackage{amssymb}
\usepackage{amsmath}
\usepackage{graphicx}
\usepackage{epsfig}
\usepackage{subfigure}
\usepackage{color}

\begin{document}

\title{$\eta $-pairing states in the Hubbard model with non-uniform Hubbard
interaction}
\author{D. K. He}
\author{Z. Song}
\email{songtc@nankai.edu.cn}
\affiliation{School of Physics, Nankai University, Tianjin 300071, China}
\begin{abstract}
The existence of $\eta $-pairing eigenstates in the fermionic Hubbard model
is fundamentally rooted in the $\eta $-pairing symmetry, which may hold for
systems with non-uniform Hubbard interaction $U$. In this work, we present a
generalized Hubbard model containing a variety of pseudo-spin terms that
break the SO$_{4}$ symmetry but retain the $\eta $-pairing symmetry. This
allows us to construct a variety of correlated systems possessing $\eta $%
-pairing eigenstates.\ We exemplify our findings by considering a modified
Hubbard model associated with alternative magnetic fields and on-site
repulsion. We find that the same quasi-$\eta $-pairing eigenstate exhibits
two distinct dynamic behaviors in the two models. Numerical results of the
time evolution driven by several typical Hamiltonians accord with the
analytic predictions and provide a way of the control of an $\eta $-pairing
wavepacket with the aid of a time-dependent Hamiltonian.
\end{abstract}

\maketitle

\section{Introduction}

New developments in quantum simulations of the Hubbard model with ultracold
atoms have given us a powerful tool to study the low-temperature properties
of strongly correlated systems \cite%
{bakr2009,parsons2015,cheuk2015,cheuk2016,parsons2016,esslinger2010}. The
Hubbard model, a fundamental minimal framework for describing interacting
fermions, is characterized by two key principles: spin-1/2 fermions can move
between lattice sites and experience on-site interactions with one another 
\cite{Hubbard63, Kanamori63, Gutzwiller63}. Remarkably, despite its
straightforward nature, this model gives rise to a diverse array of complex
quantum many-body behaviors, including magnetism and superconductivity \cite%
{Arovas21, Qin21}. By studying the properties of the Hubbard model, we gain
valuable understanding of how order can arise through the combined effects
of the energy band structure of free fermions in the external field,
density-density Coulomb interactions, and spin-spin interactions. However,
even the simplest Hubbard model, which features only nearest-neighbor
hopping and on-site interactions, rarely yields exact results. Obtaining
analytical solutions for this model is extremely difficult, except in the
one-dimensional case \cite{LiebWu68, 1dHubbard_book}.

Furtunately, a set of exact eigenstates, refered to as $\eta $-pairing
states, of the Hubbard model can be constructed through an $\eta $-pairing
operators \cite{Yang89}. The existence of $\eta $-pairing states in the
fermionic Hubbard model is fundamentally rooted in the $\eta $-pairing
symmetry of the Hubbard Hamiltonian \cite{YangZhang90, Pernici90}. One of
the most compelling findings is the potential for superconductivity, which
arises due to the off-diagonal long-range order (ODLRO) exhibited by these
states. This has led to the development of various non-equilibrium
protocols, such as external field-driven \cite{Rosch08, Kantian10, Kaneko19,
Kaneko20, Ejima20, Werner19, Li20, Kitamura16, Peronaci20, Cook20,
Tindall21, Tindall21_2, Diehl08, Kraus08, Bernier13, Buca19, Tindall19,
Tsuji21, Nakagawa21, Murakami21,zhang2022steady},
photodoping strategies \cite%
{iwai2003,rosch2008,sensarma2010,eckstein2011,ichikawa2011,lenarvcivc2013,stojchevska2014,mitrano2014,werner2019,peronaci2020,li2020}
and dissipation-based methods \cite%
{diehl2008,kraus2008,coulthard2017,werner2019,zhang2020b,zhang2021eta,yang2022dynamic}, aimed at selectively generating these
superconducting-like states. Recently, the $\eta $-pairing\ states in
systems with edge states \cite{zhang2021topologically} and on a moire
lattice \cite{wang2024flat} have also been studied.

So far, most studies have focused on systems with uniform Hubbard
interaction $U$, which is intuitively seen as a necessary condition for the
validity of the $\eta $-pairing symmetry. In this work, we propose a
generalized Hubbard model containing a variety of terms that break the SO$%
_{4}$ symmetry but retain the $\eta $-pairing symmetry. We find that the
distribution of Hubbard interaction strength can be non-uniform. There
exists a variety of Hamiltonians that exhibit $\eta $-pairing symmetry, and
then possess eigenstates with ODLRO. We exemplify our findings by
considering a modified Hubbard model associated with alternative magnetic
fields and on-site repulsion. In order to shed light on the staggered
Hubbard model with a clear physical picture, we establish an effective
Hamiltonian to examine the single-doublon dynamics in comparison with that
of the simple uniform Hubbard model. We find that the same $\eta $-pairing
eigenstate exhibits two distinct dynamic behaviors. To demonstrate this
point, we perform numerical simulations of the time evolution of an initial $%
\eta $-pairing Gaussian wavepacket driven by several typical Hamiltonians,
including one for a quench in the Hubbard model. The results indicate that
the proposed staggered Hubbard model can be utilized for the control of an $%
\eta $-pairing state with the aid of a time-dependent Hamiltonian.

The paper is organized as follows. In Sec. ~\ref{Model with eta symmetry},
we introduce a generalized Hubbard model that incorporates $\eta $-symmetry.
Sec. ~\ref{Effective Hamiltonian} focuses on the staggered Hubbard model,
where we derive its effective Hamiltonian in the large-$U$ limit. Sec. ~\ref%
{Quench dynamics} is dedicated to exploring the numerical simulations of
quench dynamics. Finally, Sec. ~\ref{Summary} provides a summary of the key
findings and conclusions of this study.

\section{Model with $\protect\eta $\ symmetry}

\label{Model with eta symmetry}

In this section, we introduce a family of tight-binding models of spinful
fermions, which possess $\eta $-pairing symmetry. These models cover the
original Hubbard model investigated in the seminal work \cite{Yang89}.
Consider a generalized tight-binding Hamiltonian of interacting fermions on
a bipartite lattice%
\begin{equation}
H=\sum_{i\in A,j\in B}^{N}\sum_{\sigma =\uparrow ,\downarrow }\kappa
_{ij}c_{i,\sigma }^{\dagger }c_{j,\sigma }+\text{\textrm{H.c.}}+\mathcal{F}%
(\left\{ s_{l}^{x},s_{l}^{y},s_{l}^{z}\right\} ),  \label{H}
\end{equation}%
where the operator $c_{j,\sigma }$ is the annihilation operator of a spin-$%
\sigma $ fermion at site $j$, satisfying the usual fermion anticommutation
relations $\{c_{i,\sigma }^{\dagger },$ $c_{j,\sigma ^{\prime }}\}=\delta
_{i,j}\delta _{\sigma ,\sigma ^{\prime }}$ and $\{c_{i,\sigma },$ $%
c_{j,\sigma ^{\prime }}\}=0$. The system consists of two sublattices, $A$
and $B$, such that each site in $A$ is only connected to sites in $B$. The
first term is the usual hopping term. In this work, the hopping matrix
elements $\kappa _{ij}$ are required to be real and satisfy $\kappa
_{ij}=\kappa _{ji}$. The second term is the focus of this work. It is
spin-dependent and is an arbitrary function of three sets of operators $%
\{s_{l}^{x},s_{l}^{y},s_{l}^{z},$ $l\in \left[ 1,N\right] \}$. For
convenience and clarity, the number of sublattice sites and the number of
filled particles are denoted by $N_{A}$($N_{B}$) and $M_{A}$($M_{B}$),
respectively. Here, the spin operators $\left\{ s_{l}^{\alpha },\alpha
=x,y,z\right\} $ for fermions at site $l$\ are defined as

\begin{eqnarray}
&&s_{l}^{+}=\left( s_{l}^{-}\right) ^{\dag }={s_{l}^{x}+is_{l}^{y}=c_{l,%
\uparrow }^{\dagger }c_{l,\downarrow },}  \notag \\
&&{s_{l}^{z}=\left( n_{l,\uparrow }-n_{l,\downarrow }\right) /2,}
\end{eqnarray}%
which obey the Lie algebra, i.e., $[s_{j}^{+},$ $s_{j}^{-}]=2s_{j}^{z}$, and 
$[s_{j}^{z},$ $s_{j}^{\pm }]=\pm s_{j}^{\pm }$, and the number operator is
given by $n_{j,\sigma }=c_{j,\sigma }^{\dagger }c_{j,\sigma }$. Note that
such spin operators cannot be directly represented by the Pauli matrices,
since the quantum numbers of the pseudo-spin operator are $0$, $0$, and $1/2$%
, respectively. For instance, we have ${s_{l}^{z}c_{l,\uparrow }^{\dagger
}c_{l,\downarrow }^{\dag }}\left\vert 0\right\rangle ={s_{l}^{z}}\left\vert
0\right\rangle =0$, while ${s_{l}^{z}c_{l,\sigma }^{\dagger }}\left\vert
0\right\rangle =\frac{1}{2}{c_{l,\sigma }^{\dagger }}\left\vert
0\right\rangle $.\ 

In general, the existence of $\eta $-pairing states in the fermionic Hubbard
model is fundamentally rooted in the $\eta $ symmetry, which relates to
spinless quasiparticles. The corresponding generators can be given as 
\begin{eqnarray}
\eta ^{+} &=&\left( \eta ^{-}\right) ^{\dagger }=\sum_{j\in A\cup B}\eta
_{j}^{+},  \notag \\
\eta ^{z} &=&\sum_{j\in A\cup B}\eta _{j}^{z},
\end{eqnarray}%
with $\eta _{j}^{+}=\lambda c_{j,\uparrow }^{\dagger }c_{j,\downarrow
}^{\dagger }$ and $\eta _{j}^{z}=\left( n_{j,\uparrow }+n_{j,\downarrow
}-1\right) /2$ satisfying the commutation relation, i.e., $[\eta _{j}^{+},$ $%
\eta _{j}^{-}]=2\eta _{j}^{z}$, and $[\eta _{j}^{z},$ $\eta _{j}^{\pm }]=\pm
\eta _{j}^{\pm }$. Here, we take $\lambda =1$ for $j\in \left\{ A\right\} $
and $\lambda =-1$ for $j\in \left\{ B\right\} $. Similarly, the quantum
numbers of the pseudo-spin operator $\eta $ are also $0$, $0$, and $1/2$,
respectively. For instance, we have $\eta _{l}^{z}{c_{l,\uparrow }^{\dagger }%
}\left\vert 0\right\rangle =\eta _{l}^{z}{c_{l,\downarrow }^{\dag }}%
\left\vert 0\right\rangle =0$, while $\eta _{l}^{z}{c_{l,\uparrow }^{\dagger
}c_{l,\downarrow }^{\dag }}\left\vert 0\right\rangle =\frac{1}{2}{%
c_{l,\uparrow }^{\dagger }c_{l,\downarrow }^{\dag }}\left\vert
0\right\rangle $ and $\eta _{l}^{z}\left\vert 0\right\rangle =-\frac{1}{2}%
\left\vert 0\right\rangle $.\ 

Importantly, the generalized Hamiltonian $H$,\ given in Eq. (\ref{H}),
possesses the $\eta $ symmetry, as evidenced by the following identities {%
\begin{equation}
\left[ H,\eta ^{\pm }\right] =\left[ H,\eta ^{z}\right] =0,
\end{equation}%
}which are based on the fundamental commutation relation [WP\_SR]%
\begin{equation}
\left[ \eta _{l}^{\alpha },s_{j}^{\beta }\right] =0,
\end{equation}%
for arbitrary $\left( l,j\right) $\ and $\left( \alpha ,\beta \right) $. The
implication of the symmerty is straightforward: $\eta ^{+}$ acts as a
generator of $\eta $-pairing eigenstates of the Hamiltonian, regardless the
details of the function $\mathcal{F}$, according to the spectrum-generating
algebras. Specifically, based on a given eigenstate of $H$, satisfying $%
H\left\vert \psi _{0}\right\rangle =E_{0}\left\vert \psi _{0}\right\rangle $%
, a set of eigenstates can be constructed by 
\begin{equation}
\left\vert \psi _{n}\right\rangle =\frac{1}{n!\sqrt{C_{L}^{n}}}\left( \eta
^{+}\right) ^{n}\left\vert \psi _{0}\right\rangle \neq 0,
\end{equation}%
satisfying 
\begin{equation}
H\left\vert \psi _{n}\right\rangle =E_{0}\left\vert \psi _{n}\right\rangle .
\end{equation}%
One of the simplest eigenstates can be the vaccum state of the fermion
operator, given by $\left\vert \psi _{0}\right\rangle =\left\vert
0\right\rangle $,\ due to the relation 
\begin{equation}
s_{l}^{\alpha }\left\vert 0\right\rangle =0,
\end{equation}%
\ for any given $l$\ and $\alpha $, where $E_{0}$\ is given by $\mathcal{F}%
\left\vert 0\right\rangle =E_{0}\left\vert 0\right\rangle $.\ We note that
such a set of eigenstates are still eigenstates of the pseudospin with $s=0$%
. However, it is not guaranteed that other $\eta $-pairing eigenstates are
also eigenstates of the pseudospin, because the term $\mathcal{F}$\ breaks
the SU(2) symmetry. Now we turn to investigate the case with explicit form
of $\mathcal{F}$, revealing the connection to the original Hubbard model.

(i) Quadratic term$\ \mathcal{F}=\mathcal{F}\left[ \left( s_{l}^{z}\right)
^{2},s_{l}^{z}\right] $. The linear term of the operator $s_{l}^{\alpha }$
is trivial because it does not contain the particle-particle interaction.
However, the quadratic term is nontrivial since we do not have $\left(
s_{l}^{z}\right) ^{2}=1/4$\ as mentioned above. Specifically, we take the
form%
\begin{equation}
\mathcal{F}=-\sum_{l\in A\cup B}\frac{U_{l}}{2}[4\left( s_{l}^{z}\right)
^{2}+2\left( \lambda _{l}-1\right) s_{l}^{z}],
\end{equation}%
where $\left\{ \lambda _{l}\right\} $\ is an arbitrary set of numbers,
including complex numbers. This can be expressed in the from%
\begin{equation}
\mathcal{F}=\sum_{l\in A\cup B}U_{l}(n_{l,\uparrow }n_{l,\downarrow }-\frac{%
\lambda _{l}}{2}n_{l,\uparrow }-\frac{2-\lambda _{l}}{2}n_{l,\downarrow }).
\end{equation}%
It has the following implications. (a) When taking $U_{l}=U$ and $\lambda
_{l}=1$, the term $\mathcal{F}$ reduces to 
\begin{equation}
U\sum_{l\in A\cup B}[(n_{l,\uparrow }-\frac{1}{2})(n_{l,\downarrow }-\frac{1%
}{2})-\frac{1}{4}],
\end{equation}%
which is the original Hubbard interaction term in the previous works \cite%
{Yang89}. In this case, we have $\left[ \mathcal{F},s_{j}^{\beta }\right] =%
\left[ \mathcal{F},\eta _{l}^{\alpha }\right] =0$, that is, the system has SO%
$_{4}$ symmetry. Then we can have $N_{A}\neq N_{B}$ and $M_{A}\neq M_{B}$.\
(b) In contrast, the term $\mathcal{F}$ with $\lambda _{l}\neq 1/2$ spoils
the SU(2) symmetry but retains $\eta $ symmetry. (c) Although the paramters $%
\left\{ U_{l},\lambda _{l}\right\} $ are arbitrary, we always have $\mathcal{%
F}{c_{l,\uparrow }^{\dagger }c_{l,\downarrow }^{\dag }}\left\vert
0\right\rangle =\mathcal{F}\left\vert 0\right\rangle =0$.\ This sheds light
on the underlying mechanism of the existence of $\eta $-pairing eigenstates.

(ii) Higher-order terms. Now we turn to the case where the term $\mathcal{F}$
contains higher-order terms of the operator $s_{l}^{\alpha }$. In the
following, we will show that a higher-order term can always be reduced to $%
\mathcal{F}=\mathcal{F}\left[ \left( s_{l}^{z}\right) ^{2},s_{l}^{z}\right] $%
. In fact, based on the identies%
\begin{equation}
\left( s^{z}\right) ^{3}=\frac{1}{4}s^{z},\left( s^{z}\right) ^{4}=\frac{1}{4%
}\left( s^{z}\right) ^{2},
\end{equation}%
we obtain%
\begin{equation}
\left( s^{z}\right) ^{2n-1}=\left( \frac{1}{4}\right) ^{n-1}\left(
s^{z}\right) ,
\end{equation}%
and%
\begin{equation}
\left( s^{z}\right) ^{2n}=\left( \frac{1}{4}\right) ^{n-1}\left(
s^{z}\right) ^{2},
\end{equation}%
for any given $n$. Then we conclude that the relations 
\begin{equation}
\mathcal{F}\left[ \left( s_{l}^{z}\right) ^{n}\right] {c_{l,\uparrow
}^{\dagger }c_{l,\downarrow }^{\dag }}\left\vert 0\right\rangle =\mathcal{F}%
\left[ \left( s_{l}^{z}\right) ^{n}\right] \left\vert 0\right\rangle =0,
\end{equation}%
still hold. Moreover, the term $\mathcal{F}(\left\{
s_{l}^{x},s_{l}^{y},s_{l}^{z}\right\} )$\ can also represent two-site and
multi-site couplings, such as Heisenberg, Dzyaloshinskii-Moriya (DM)
interaction, and other few-body dipole-dipole interactions \cite%
{zhang2015topological,zhang2017majorana,vafek2017entanglement,spielman2024quantum}%
.

\section{Effective Hamiltonian}

\label{Effective Hamiltonian}

\begin{figure*}[th]
\centering
\includegraphics[width=0.9\textwidth]{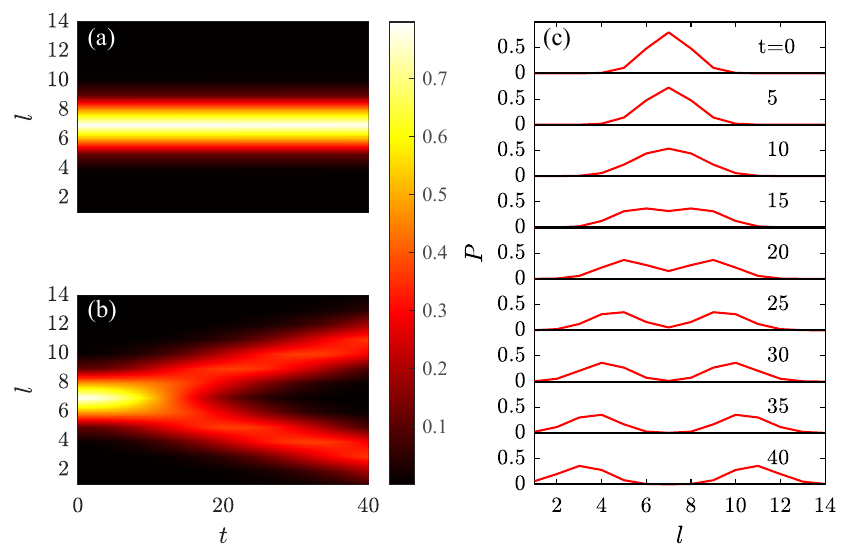}
\caption{Profiles of evolved wavepackets for initial states expressed as $|%
\protect\psi (0)\rangle $\ in Eq. (\protect\ref{initial state_m}) with $m=1$
and $\protect\alpha =0.5$, driven by the Hamitonians $H_{\text{\textrm{Unif}}%
}$ and $H_{\text{\textrm{Stag}}}$, respectively, with $U=5,\protect\kappa %
=0.1$. The color contour plot of the Dirac norm $P\left( l,t\right) $,
defined in Eq. (\protect\ref{P(l,t)}), obtained by numerical simulations for
the Hamiltonians $H_{\text{\textrm{Unif}}}$ and $H_{\text{\textrm{Stag}}}$,
is shown in (a) and (b), respectively. (c) Profiles of evolved states under
the Hamiltonian $H_{\text{\textrm{Stag}}}$, at several instants, obtained by
numerical simulations. These results show that the two evolved states
exhibit distinct behaviors, as we predicted. The state in (a) remains
unchanged, while the state in (b) splits into two wavepackets.}
\label{fig1}
\end{figure*}
In this section, we aim to shed light on this issue with a clear physical
picture. To achieve this, we investigate two representative Hamiltonians:
one featuring a uniform structure and the other featuring a staggered
structure. We will examine the single-doublon dynamics in both systems and
highlight the differences between them.

(i) The uniform Hubbard chain. The corresponding Hamiltonian is

\begin{eqnarray}
&&H_{\text{\textrm{Unif}}}=\kappa \sum_{l=1}^{2N}[\sum_{\sigma =\uparrow
,\downarrow }c_{l,\sigma }^{\dagger }c_{l+1,\sigma }+\text{\textrm{H.c.}} 
\notag \\
&&+U(n_{l,\uparrow }-\frac{1}{2})(n_{l,\downarrow }-\frac{1}{2})-\frac{1}{4}%
],
\end{eqnarray}%
where the corresponding term $\mathcal{F}$ is obtained by taking $U_{l}=U$, $%
\kappa _{l(l+1)}=\kappa $ and $\lambda _{l}=1$, respectively. It has been
shown that in the strong correlation limit with $\left\vert U\right\vert \gg
\kappa $, the dynamics of the doublon can be described by the following
effective Hamiltonian%
\begin{equation}
H_{\text{\textrm{Unif}}}^{\mathrm{eff}}=\frac{4\kappa ^{2}}{U}%
\sum_{l=1}^{2N}\left( \mathbf{\eta }_{l}\cdot \mathbf{\eta }_{l+1}-\frac{1}{4%
}\right) ,
\end{equation}%
which is an isotropic Heisenberg for a pseudo-spin chain. The derivation of
the Hamiltonian $H_{\text{\textrm{Unif}}}^{\mathrm{eff}}$\ is given in the
Ref. \cite{zhang2021eta}. We are interested in the single-doublon dynamics,
which is governed by the effective Hamiltonian%
\begin{equation}
h_{\text{\textrm{Unif}}}=\frac{2\kappa ^{2}}{U}\sum_{j=1}^{2N}(\left\vert
2j\right\rangle \left\langle 2j+2\right\vert +\text{\textrm{H.c.}}),
\label{h_Unif}
\end{equation}%
where the basis set $\left\{ \left\vert l\right\rangle ,l\in \left[ 1,4N%
\right] \right\} $ is given by%
\begin{equation}
\left\vert 2j\right\rangle ={c_{2j,\uparrow }^{\dagger }c_{2j,\downarrow
}^{\dag }}\left\vert 0\right\rangle ,
\end{equation}%
and%
\begin{equation}
\left\vert 2j-1\right\rangle ={c_{2j-1,\uparrow }^{\dagger }c_{2j,\downarrow
}^{\dag }}\left( {c_{2j-2,\downarrow }^{\dag }c_{2j-1,\uparrow }^{\dagger }}%
\right) \left\vert 0\right\rangle .
\end{equation}%
Obviously, it describes a single-particle chain with uniform NN hopping with
strength $2\kappa ^{2}/U$. The dispersion of the single-doublon is $4\kappa
^{2}/U\cos k$, where the wave vector is given by $k=2\pi n/(2N)$\ ($n\in %
\left[ 1,2N\right] $). In this framework, the $\eta $-pairing state
corresponds to the plane wave with $k=\pi $.\ Importantly, the
single-doublon wavepacket near the $\eta $-pairing state\ has zero group
velocity. In this sense, the condensate $\eta $-pairing states are frozen
states.

\begin{figure*}[th]
\centering
\includegraphics[width=0.9\textwidth]{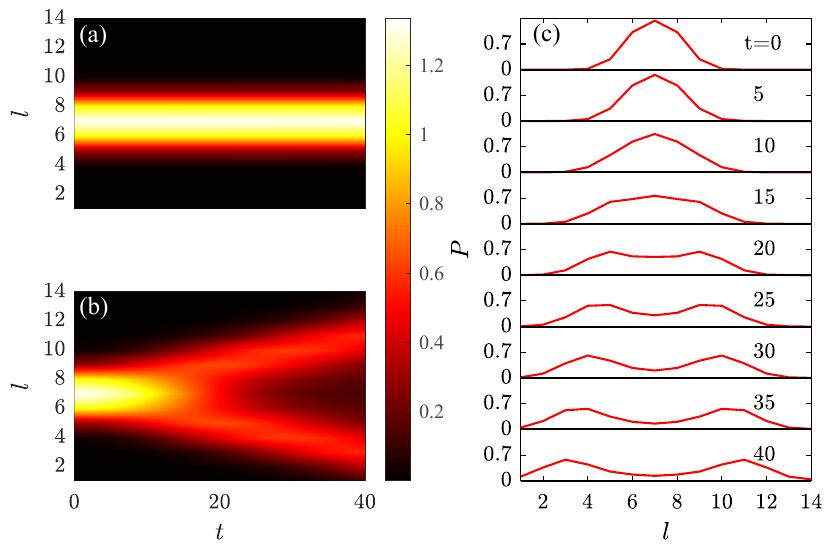}
\caption{The same plots as that in Fig. \protect\ref{fig1}, but for initial
states expressed as $|\protect\psi (0)\rangle $\ in Eq. (\protect\ref%
{initial state_m}) with $m=2$. These results show that the two evolved
states exhibit similar behaviors as the case with $m=1$. The state in (a)
remains unchanged, while the state in (b) splits into two wavepackets.}
\label{fig2}
\end{figure*}
(ii) The staggered Hubbard chain. The corresponding Hamiltonian is 
\begin{eqnarray}
&&H_{\text{\textrm{Stag}}}=\kappa \sum_{l=1}^{2N}\sum_{\sigma =\uparrow
,\downarrow }c_{l,\sigma }^{\dagger }c_{l+1,\sigma }+\text{\textrm{H.c.}}) 
\notag \\
&&+U\sum_{l\in \mathrm{odd}}(n_{l,\uparrow }-\frac{1}{2})(n_{l,\downarrow }-%
\frac{1}{2})  \notag \\
&&+U\sum_{l\in \mathrm{even}}(n_{l,\uparrow }+\frac{1}{2})(n_{l,\downarrow }-%
\frac{3}{2})+NU,
\end{eqnarray}%
where the corresponding term $\mathcal{F}$ is obtained by taking $U_{l}=U$, $%
\kappa _{l(l+1)}=\kappa $ and $\lambda _{l}=1$ for odd $l$ and $\lambda
_{l}=3$ even $l$. It is a little difficult to establish an effective
Hamiltonian for many-doublon dynamics in strong correlation limit with $%
\left\vert U\right\vert \gg \kappa $, because there are many single-occupied
configurations in the same energy shell of the doublons. In fact, we always
have 
\begin{eqnarray}
&&H_{\text{\textrm{Stag}}}\left( 0\right) c_{j,\uparrow }^{\dagger
}c_{j,\downarrow }^{\dagger }|0\rangle =H_{\text{\textrm{Stag}}}\left(
0\right) c_{j+1,\downarrow }^{\dagger }c_{j+1,\uparrow }^{\dagger }|0\rangle 
\notag \\
&=&H_{\text{\textrm{Stag}}}\left( 0\right) c_{j,\uparrow }^{\dagger
}c_{j+1,\downarrow }^{\dagger }|0\rangle =H_{\text{\textrm{Stag}}}\left(
0\right) c_{j-1,\downarrow }^{\dagger }c_{j,\uparrow }^{\dagger }|0\rangle  
\notag \\
&=&0,
\end{eqnarray}%
for any odd $j$ and the Hamiltonian $H_{\text{\textrm{Stag}}}\left( 0\right) 
$\ denotes the Hamiltonian at zero $\kappa $. We are interested in the
single-doublon dynamics, which is governed by the effective Hamiltonian%
\begin{eqnarray}
h_{\text{\textrm{Stag}}} &=&\sum_{l=1}^{4N}[(\kappa \left\vert
l\right\rangle \left\langle l+1\right\vert +\frac{\kappa ^{2}}{2U}\left\vert
l\right\rangle \left\langle l+2\right\vert +\text{\textrm{H.c.}})  \notag \\
&&+\frac{\kappa ^{2}}{4U}\left( 1-\left( -1\right) ^{l}\right) \left\vert
l\right\rangle \left\langle l\right\vert ].  \label{h_Stag}
\end{eqnarray}%
It describes a single-particle chain with uniform NN hopping with strength $%
\kappa $ in the large $U$\ limit. The dispersion of the single-doublon is $%
2\kappa \cos k$, where the wave vector is given by $k=2\pi n/(4N)$\ ($n\in %
\left[ 1,4N\right] $). In this framework, the $\eta $-pairing state
corresponds to the superposition of two plane waves with $k=\pm \pi /2$.\
Importantly, the single-doublon wavepacket near the $\eta $-pairing state\
has maximal group velocity. In this sense, the condensate $\eta $-pairing
states are mobile states.

When comparing the single-doublon effective Hamiltonians for cases (i) and
(ii), the underlying physical picture becomes quite clear. The behavior of
the doublon can be effectively modeled by a noninteracting hardcore boson
system on a chain lattice, incorporating both nearest-neighbor (NN) and
next-nearest-neighbor (NNN) hopping terms. Specifically, case (i) features a
small but nonzero hopping strength across even sites, whereas case (ii) is
characterized by a large and nonzero hopping strength across any
nearest-neighbor sites. This framework is significant because it sheds light
on the transition between two distinct $\eta $-pairing dynamics, which are
driven by the external field.

\section{Quench dynamics}

\label{Quench dynamics}

In this section, we will demonstrate the above results through a quench
process. First, we investigate this problem analytically based on the
effective Hamiltonian obtained in the last section. Then, we perform
numerical simulations for the time evolution of an initial $\eta $-pairing
Gaussian wavepacket driven by several typical Hamiltonians, including one
for a quench in the Hubbard model.

\begin{figure*}[th]
\centering
\includegraphics[width=0.95\textwidth]{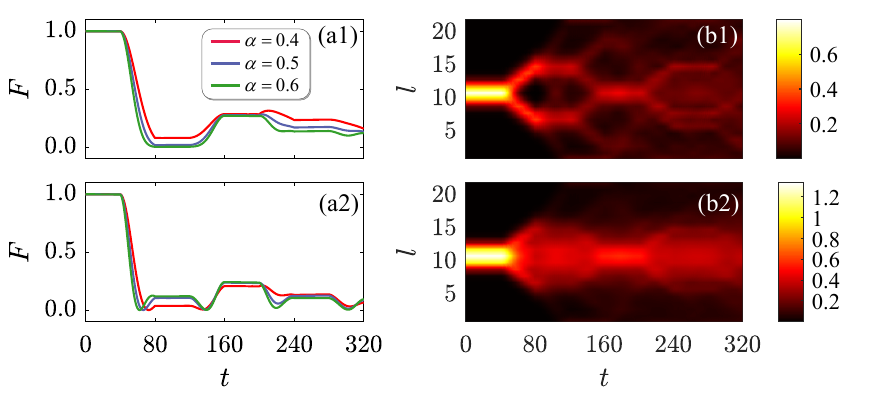}
\caption{Plots of the fidelity $F\left( t\right) $, given in Eq. (\protect
\ref{Ft}) and the Dirac norm $P\left( l,t\right) $, given in Eq. (\protect
\ref{P(l,t)}) of the initial state $\left\vert \protect\psi \left( 0\right)
\right\rangle $, given in Eq. (\protect\ref{initial state_m}) under the
driven Hamiltonian $H_{\mathrm{quen}}\left( t\right) $, given in Eq. (%
\protect\ref{H_quen}). In plots (a1) and (a2), the fidelity is calculated as
a function of time for the initial state $\left\vert \protect\psi \left(
0\right) \right\rangle $ with three representative values of $\protect\alpha 
$, indicated in the panel. Specifically, (a1) shows the results for $m=1$,
while (a2) shows the results for $m=2$. Plots (b1) and (b2) display the
color contour maps of $P\left( l,t\right) $ for the initial state $a=0.5$.
Here, (b1) corresponds to $m=1$, and (b2) corresponds to $m=2$. In all
cases, the parameters are kept constant as $U=10$, $\protect\kappa =0.1$, $2%
\protect\tau =80$, and $N=22$. We find from the plots (a1) and (a2) that in
each interval of time $2n\protect\tau \leqslant t<\left( 2n+1\right) \protect%
\tau $ ($n=0,1,2,3$), the fidelity exhibits a plateau, indicating the frozen 
$\protect\eta $-pairing states. The result is not sensitive to the shape or
the number of particles in the initial state. The plots (b1) and (b2) offer
a visual illustration of the fidelity. They clearly show that the
enhancement of the third plateau is due to the interference between two
sub-wavepackets.}
\label{fig3}
\end{figure*}

We start by considering the time evolution of an initial state%
\begin{equation}
|\psi (0)\rangle =\frac{1}{\sqrt{\Omega }}\sum_{l}e^{i\pi l}e^{-\alpha
^{2}(2l-2l_{c})^{2}}\left\vert 2l\right\rangle ,
\end{equation}%
where $\Omega $\ is the normalization factor. The state $|\psi (0)\rangle $\
represents a Gaussian wavepacket of a doublon with its center at site $%
2l_{c} $. The width of the wavepacket is determined by the factor $\alpha $.
When $\alpha $ vanishes, the state $|\psi (0)\rangle $\ becomes the $\eta $%
-pairing eigenstates of $h_{\text{\textrm{Unif}}}$\ and\ $h_{\text{\textrm{%
Stag}}}$. For finite $\alpha $, the corresponding evoleved states under the
Hamiltonians $h_{\text{\textrm{Unif}}}$\ and\ $h_{\text{\textrm{Stag}}}$ are%
\begin{equation}
|\psi _{\mathrm{A}}(t)\rangle =e^{-ih_{\text{\textrm{Unif}}}t}|\psi
(0)\rangle ,
\end{equation}%
and%
\begin{equation}
|\psi _{\mathrm{B}}(t)\rangle =e^{-ih_{\text{\textrm{Stag}}}t}|\psi
(0)\rangle ,
\end{equation}%
respectively.

In a short time scale $t\sim 1/\kappa $, we have

\begin{equation}
|\psi _{\mathrm{A}}(t)\rangle \approx |\psi (0)\rangle ,
\end{equation}%
due to the vanishing group velocity of the wavepacket. In contrast, we have%
\begin{equation}
|\psi _{\mathrm{B}}(t)\rangle \approx |\phi _{+}(t)\rangle +|\phi
_{-}(t)\rangle ,
\end{equation}%
where the evolved states%
\begin{equation}
|\phi _{\pm }(t)\rangle \sim \sum_{l}e^{i\pi l/2}e^{-\alpha ^{2}(l-2l_{c}\mp
v_{g}t)^{2}}\left\vert l\right\rangle ,
\end{equation}%
represent two moving wavepackets with group velocity $v_{g}=2\kappa $. This
indicates that the same initial state should exhibit distinct dynamical
behaviors. This conclusion should also hold true for many-doublon initial
states.

To demonstrate and verify the conclusion, numerical simulations are
performed for the time evolution of the initial state in the form

\begin{equation}
|\psi (0)\rangle =\frac{1}{\sqrt{\Lambda }}\left( \sum_{l}e^{i\pi
l}e^{-\alpha ^{2}(l-l_{c})^{2}}{c_{l,\uparrow }^{\dagger }c_{l,\downarrow
}^{\dag }}\right) ^{m}\left\vert 0\right\rangle ,  \label{initial state_m}
\end{equation}%
with small $m$, where $\sqrt{\Lambda }$\ is the normalization factor.\ It
represents $m$-pair condensate state. The driven Hamiltonian is an
alternating quench Hamiltonian with $H_{\text{\textrm{Unif}}}$, $H_{\text{%
\textrm{Stag}}}$, and $H_{\text{\textrm{quen}}}(t)$, defined as

\begin{equation}
H_{\mathrm{quen}}(t)=\left\{ 
\begin{array}{cc}
H_{\text{\textrm{Unif}}}, & 2n\tau \leqslant t<\left( 2n+1\right) \tau \\ 
H_{\text{\textrm{Stag}}}, & \left( 2n+1\right) \tau \leqslant t<\left(
2n+2\right) \tau%
\end{array}%
\right. ,  \label{H_quen}
\end{equation}%
with $n=0,1,2,...$. We note that the Hamiltonian $H_{\mathrm{quen}}(t)$\ is
periodic with period $2\tau $.

To characterize the evolved state, {we will compute the Dirac probability
distribution in real space} 
\begin{equation}
P(l,t)=\left\vert {c_{l,\uparrow }}e^{-iHt}|\psi (0)\rangle \right\vert
^{2}+\left\vert {c_{l,\downarrow }}e^{-iHt}|\psi (0)\rangle \right\vert ^{2},
\label{P(l,t)}
\end{equation}%
and the fidelity

\begin{equation}
F(t)=\frac{\left\vert \left\langle \psi (0)\right\vert e^{-iHt}|\psi
(0)\rangle \right\vert ^{2}}{\left\vert e^{-iHt}|\psi (0)\rangle \right\vert
^{2}},  \label{Ft}
\end{equation}%
{for the time evolution driven by the three types of Hamiltonians (}${H}${: }%
$H_{\text{\textrm{Unif}}}$, $H_{\text{\textrm{Stag}}}$, $H_{\text{\textrm{%
quen}}}${). }For time-dependent simulations, the time-evolution operator can
be approximated using the Trotter-Suzuki decomposition \cite%
{blanes1998magnus}, which splits the Hamiltonian into simpler parts and
applies them sequentially.

The numerical results for the cases with the Hamiltonians $H_{\text{\textrm{%
Unif}}}$ and $H_{\text{\textrm{Stag}}}$\ are plotted in Fig. \ref{fig1} and %
\ref{fig2}, respectively. As we predicted, the dynamics of both single and
double $\eta $-pair Gaussian wavepacket initial states are almost the same.
They almost remain unchanged in the presence of the Hamiltonian $H_{\text{%
\textrm{Unif}}}$, due to their vanisihing group velocities. In contrast,
they split into two sub-wavepackets with opposite constant group velocities
in the presence of the Hamiltonian $H_{\text{\textrm{Stag}}}$.$\ $For the
case with $H_{\mathrm{quen}}$, the result is plotted Fig. \ref{fig3}. The
obtained results also accord with our predictions obtained from the
effective Hamiltonians. {Therefore, we conclude that the same }$\eta ${%
-pairing state exhibits distinct dynamics in the two Hamiltonians }$H_{\text{%
\textrm{Unif}}}${\ and }$H_{\text{\textrm{Stag}}}${. This enables the
control of an }$\eta${-pairing state with the aid of a time-dependent
Hamiltonian.}

Before concluding this section, we intend to explore the nature of the split
sub-wavepackets via the subsequent analytical examinations. We define the
operators $\zeta _{\pm }${}as follows%
\begin{eqnarray}
\zeta _{\pm } &=&\sum_{l=1}^{N}(-{c_{2l-1,\uparrow }^{\dagger
}c_{2l-1,\downarrow }^{\dag }+{c_{2l,\uparrow }^{\dagger }c_{2l,\downarrow
}^{\dag }}}  \notag \\
&&\mp ic_{2l-1,\uparrow }^{\dagger }c_{2l,\downarrow }^{\dag }\pm
ic_{2l+1,\uparrow }^{\dagger }c_{2l,\downarrow }^{\dag }).
\end{eqnarray}%
Using these operators, the $\eta $-operator can be written as%
\begin{equation}
\eta ^{+}=\zeta _{+}+\zeta _{-}.
\end{equation}%
It is evident that, unlike the operator $\eta ^{+}$, the operators $\zeta
_{\pm }$ do not commute with {two Hamiltonians }$H_{\text{\textrm{Unif}}}${\
and }$H_{\text{\textrm{Stag}}}$. We observe that the states $\zeta
_{+}\left\vert 0\right\rangle $ and $\zeta _{-}\left\vert 0\right\rangle $
are eigenstates of $H_{\text{\textrm{Stag}}}$ in large $U$ limit. However,
the multi-pair condensate states $\left( \zeta _{+}\right) ^{n}\left\vert
0\right\rangle $ and $\left( \zeta _{-}\right) ^{n}\left\vert 0\right\rangle 
$ ($n>1$) are not eigenstates. The common eigenstate $\left( \eta
^{+}\right) ^{n}\left\vert 0\right\rangle $ of the Hamiltonians $H_{\text{%
\textrm{Unif}}}$ and $H_{\text{\textrm{Stag}}}$\ can be expanded as%
\begin{equation}
\left( \eta ^{+}\right) ^{n}\left\vert 0\right\rangle
=\sum_{m=0}^{n}C_{n}^{m}\left( \zeta _{-}\right) ^{m}\left( \zeta
_{+}\right) ^{n-m}\left\vert 0\right\rangle ,
\end{equation}%
which is known to exhibit ODLRO. We aim to demonstrate that each component
state $\left\vert \phi _{n}^{\pm }\right\rangle =\left( \zeta _{\pm }\right)
^{n}\left\vert 0\right\rangle $ also possesses ODLRO. To this end, we
utilize the correlation function%
\begin{equation}
\,\mathcal{C}_{j,r}^{\pm }=\frac{\left\langle \phi _{n}^{\pm }|\eta
_{j}^{+}\eta _{j+r}|\phi _{n}^{\pm }\right\rangle }{\left\langle \phi
_{n}^{\pm }|\phi _{n}^{\pm }\right\rangle },
\end{equation}%
which serves as a measure of the ODLRO in the state $\left\vert \phi
_{n}^{\pm }\right\rangle $. Direct algebra can estimate the lower bound of
the correlation function, i.e., 
\begin{equation}
\,\mathcal{C}_{j,r}^{\pm }>\frac{%
\sum_{n_{1}=0}^{n-1}C_{2N-n_{1}-3}^{n_{1}}C_{2N-2n_{1}-3}^{n-1-n_{1}}}{%
\sum_{n_{1}=0}^{n}C_{2N-n_{1}+1}^{n_{1}}C_{2N-2n_{1}+1}^{n-n_{1}}},
\end{equation}%
which is independent of $j$\ and $r$. Although the analytic expression of
this lower bound cannot be given, numerical simulations indicate that it
remains finite for finite $n/N$. This finding suggests the presence of ODLRO
in the state $\left\vert \phi _{n}^{\pm }\right\rangle $.

\section*{Summary}

\label{Summary} In summary, we have extended the family of Hamiltonians with 
$\eta $-pairing symmetry from the simplest Hubbard model with to a variety
of correlated systems. A uniform $U$ is not a necessary condition for a
model with $\eta $-pairing symmetry. This allows us to construct a variety
of correlated systems possessing $\eta $-pairing eigenstates.\ To
demonstrate this point, we have presented a modified Hubbard model
associated with alternative magnetic fields and on-site repulsion. In
comparison with the simplest Hubbard model, we investigated the formation
mechanism of $\eta $-pairing eigenstates in the staggered Hubbard model
using single-pair effective Hamiltonians. We found that the $\eta$-pairing
eigenstate corresponds to a plane wave with $k=\pi $ for the simplest
Hubbard model, while it corresponds to a standing wave composed of two plane
waves with $k=\pm \pi /2$ for the staggered Hubbard model. This indicates
that the same quasi-$\eta $-pairing eigenstate exhibits two distinct dynamic
behaviors in the two models. This has been demonstrated by numerical
simulations of the time evolution of an $\eta $-pairing wavepacket driven by
several typical Hamiltonians. Our findings open avenues for the study of the 
$\eta $-pairing state.

\section*{Acknowledgment}

We acknowledge the support of NSFC (Grants No. 12374461).


\begin{thebibliography}{10}

\bibitem{bakr2009}
Waseem~S Bakr, Jonathon~I Gillen, Amy Peng, Simon F{\"o}lling, and Markus
  Greiner.
\newblock A quantum gas microscope for detecting single atoms in a
  hubbard-regime optical lattice.
\newblock {\em Nature}, 462(7269):74--77, 2009.

\bibitem{parsons2015}
Maxwell~F Parsons, Florian Huber, Anton Mazurenko, Christie~S Chiu, Widagdo
  Setiawan, Katherine Wooley-Brown, Sebastian Blatt, and Markus Greiner.
\newblock Site-resolved imaging of fermionic li 6 in an optical lattice.
\newblock {\em Physical review letters}, 114(21):213002, 2015.

\bibitem{cheuk2015}
Lawrence~W Cheuk, Matthew~A Nichols, Melih Okan, Thomas Gersdorf, Vinay~V
  Ramasesh, Waseem~S Bakr, Thomas Lompe, and Martin~W Zwierlein.
\newblock Quantum-gas microscope for fermionic atoms.
\newblock {\em Physical review letters}, 114(19):193001, 2015.

\bibitem{cheuk2016}
Lawrence~W Cheuk, Matthew~A Nichols, Katherine~R Lawrence, Melih Okan, Hao
  Zhang, and Martin~W Zwierlein.
\newblock Observation of 2d fermionic mott insulators of k 40 with single-site
  resolution.
\newblock {\em Physical review letters}, 116(23):235301, 2016.

\bibitem{parsons2016}
Maxwell~F Parsons, Anton Mazurenko, Christie~S Chiu, Geoffrey Ji, Daniel Greif,
  and Markus Greiner.
\newblock Site-resolved measurement of the spin-correlation function in the
  fermi-hubbard model.
\newblock {\em Science}, 353(6305):1253--1256, 2016.

\bibitem{esslinger2010}
Tilman Esslinger.
\newblock Fermi-hubbard physics with atoms in an optical lattice.
\newblock {\em Annu. Rev. Condens. Matter Phys.}, 1(1):129--152, 2010.

\bibitem{Hubbard63}
J.~Hubbard.
\newblock Electron correlations in narrow energy bands.
\newblock {\em Proc. R. Soc. Lond. A}, 276:238, 1963.

\bibitem{Kanamori63}
J.~Kanamori.
\newblock Electron correlation and ferromagnetism of transition metals.
\newblock {\em Prog. Theor. Phys.}, 30:275, 1963.

\bibitem{Gutzwiller63}
M.~C. Gutzwiller.
\newblock Effect of correlation on the ferromagnetism of transition metals.
\newblock {\em Phys. Rev. Lett.}, 10:159, 1963.

\bibitem{Arovas21}
D.~P. Arovas, E.~Berg, S.~A. Kivelson, and S.~Raghu.
\newblock The hubbard model.
\newblock {\em Annu. Rev. Condens. Matter Phys.}, 13:239, 2022.

\bibitem{Qin21}
M.~Qin, T.~Schäfer, S.~Andergassen, P.~Corboz, and E.~Gull.
\newblock The hubbard model: A computational perspective.
\newblock {\em Annu. Rev. Condens. Matter Phys.}, 13:275, 2022.

\bibitem{LiebWu68}
E.~H. Lieb and F.~Y. Wu.
\newblock Absence of mott transition in an exact solution of the short-range,
  one-band model in one dimension.
\newblock {\em Phys. Rev. Lett.}, 20:1445, 1968.

\bibitem{1dHubbard_book}
F.~H.~L. Essler, H.~Frahm, F.~Göhmann, A.~Klümper, and V.~E. Korepin.
\newblock {\em The One-Dimensional Hubbard Model}.
\newblock Cambridge University Press, Cambridge, 2010.

\bibitem{Yang89}
C.~N. Yang.
\newblock $\ensuremath{\eta}$ pairing and off-diagonal long-range order in a
  hubbard model.
\newblock {\em Phys. Rev. Lett.}, 63:2144, 1989.

\bibitem{YangZhang90}
C.~N. Yang and S.~C. Zhang.
\newblock $\mathrm{SO}4$ symmetry in a hubbard model.
\newblock {\em Mod. Phys. Lett. B}, 04:759, 1990.

\bibitem{Pernici90}
M.~Pernici.
\newblock Spin and pairing algebras and odlro in a hubbard model.
\newblock {\em Europhys. Lett.}, 12:75, 1990.

\bibitem{Rosch08}
A.~Rosch, D.~Rasch, B.~Binz, and M.~Vojta.
\newblock Metastable superfluidity of repulsive fermionic atoms in optical
  lattices.
\newblock {\em Phys. Rev. Lett.}, 101:265301, 2008.

\bibitem{Kantian10}
A.~Kantian, A.~J. Daley, and P.~Zoller.
\newblock $\ensuremath{\eta}$ condensate of fermionic atom pairs via adiabatic
  state preparation.
\newblock {\em Phys. Rev. Lett.}, 104:240406, 2010.

\bibitem{Kaneko19}
T.~Kaneko, T.~Shirakawa, S.~Sorella, and S.~Yunoki.
\newblock Photoinduced $\ensuremath{\eta}$ pairing in the hubbard model.
\newblock {\em Phys. Rev. Lett.}, 122:077002, 2019.

\bibitem{Kaneko20}
T.~Kaneko, S.~Yunoki, and A.~J. Millis.
\newblock Charge stiffness and long-range correlation in the optically induced
  $\ensuremath{\eta}$-pairing state of the one-dimensional hubbard model.
\newblock {\em Phys. Rev. Research}, 2:032027, 2020.

\bibitem{Ejima20}
S.~Ejima, T.~Kaneko, F.~Lange, S.~Yunoki, and H.~Fehske.
\newblock Photoinduced $\ensuremath{\eta}$-pairing at finite temperatures.
\newblock {\em Phys. Rev. Research}, 2:032008, 2020.

\bibitem{Werner19}
P.~Werner, J.~Li, D.~Gole\v{z}, and M.~Eckstein.
\newblock Entropy-cooled nonequilibrium states of the hubbard model.
\newblock {\em Phys. Rev. B}, 100:155130, 2019.

\bibitem{Li20}
J.~Li, D.~Golez, P.~Werner, and M.~Eckstein.
\newblock $\ensuremath{\eta}$-paired superconducting hidden phase in photodoped
  mott insulators.
\newblock {\em Phys. Rev. B}, 102:165136, 2020.

\bibitem{Kitamura16}
S.~Kitamura and H.~Aoki.
\newblock $\ensuremath{\eta}$-pairing superfluid in periodically-driven
  fermionic hubbard model with strong attraction.
\newblock {\em Phys. Rev. B}, 94:174503, 2016.

\bibitem{Peronaci20}
F.~Peronaci, O.~Parcollet, and M.~Schir\'o.
\newblock Enhancement of local pairing correlations in periodically driven mott
  insulators.
\newblock {\em Phys. Rev. B}, 101:161101, 2020.

\bibitem{Cook20}
M.~W. Cook and S.~R. Clark.
\newblock Controllable finite-momenta dynamical quasicondensation in the
  periodically driven one-dimensional fermi-hubbard model.
\newblock {\em Phys. Rev. A}, 101:033604, 2020.

\bibitem{Tindall21}
J.~Tindall, B.~Bu\v{c}a, F.~Schlawin, M.~A. Sentef, and D.~Jaksch.
\newblock Analytical solution for the steady states of the driven hubbard
  model.
\newblock {\em Phys. Rev. B}, 103:035146, 2021.

\bibitem{Tindall21_2}
J.~Tindall, B.~Bu\v{c}a, F.~Schlawin, M.~A. Sentef, and D.~Jaksch.
\newblock Lieb's {T}heorem and {M}aximum {E}ntropy {C}ondensates.
\newblock {\em Quantum}, 5:610, 2021.

\bibitem{Diehl08}
S.~Diehl, A.~Micheli, A.~Kantian, B.~Kraus, H.~P. Büchler, and P.~Zoller.
\newblock Quantum states and phases in driven open quantum systems with cold
  atoms.
\newblock {\em Nat. Phys.}, 4:878, 2008.

\bibitem{Kraus08}
B.~Kraus, H.~P. Büchler, S.~Diehl, A.~Kantian, A.~Micheli, and P.~Zoller.
\newblock Preparation of entangled states by quantum markov processes.
\newblock {\em Phys. Rev. A}, 78:042307, 2008.

\bibitem{Bernier13}
J.-S. Bernier, P.~Barmettler, D.~Poletti, and C.~Kollath.
\newblock Emergence of spatially extended pair coherence through incoherent
  local environmental coupling.
\newblock {\em Phys. Rev. A}, 87:063608, 2013.

\bibitem{Buca19}
B.~Bu\v{c}a, J.~Tindall, and D.~Jaksch.
\newblock Non-stationary coherent quantum many-body dynamics through
  dissipation.
\newblock {\em Nat. Commun.}, 10:1730, 2019.

\bibitem{Tindall19}
J.~Tindall, B.~Bu\v{c}a, J.~R. Coulthard, and D.~Jaksch.
\newblock Heating-induced long-range $\ensuremath{\eta}$ pairing in the hubbard
  model.
\newblock {\em Phys. Rev. Lett.}, 123:030603, 2019.

\bibitem{Tsuji21}
N.~Tsuji, M.~Nakagawa, and M.~Ueda.
\newblock Tachyonic and plasma instabilities of $\eta$-pairing states coupled
  to electromagnetic fields.
\newblock {\em arXiv:2103.01547}, 2021.

\bibitem{Nakagawa21}
M.~Nakagawa, N.~Tsuji, N.~Kawakami, and M.~Ueda.
\newblock $\eta$ pairing of light-emitting fermions: Nonequilibrium pairing
  mechanism at high temperatures.
\newblock {\em arXiv:2103.13624}, 2021.

\bibitem{Murakami21}
Y.~Murakami, S.~Takayoshi, T.~Kaneko, Z.~Sun, D.~Gole\v{z}, A.~J. Millis, and
  P.~Werner.
\newblock Exploring nonequilibrium phases of photo-doped mott insulators with
  generalized gibbs ensembles.
\newblock {\em Commun. Phys.}, 5:23, 2022.

\bibitem{zhang2022steady}
XZ~Zhang and Z~Song.
\newblock Steady off-diagonal long-range order state in a half-filled dimerized
  hubbard chain induced by a resonant pulsed field.
\newblock {\em Physical Review B}, 106(9):094301, 2022.

\bibitem{iwai2003}
S~Iwai, M~Ono, A~Maeda, H~Matsuzaki, H~Kishida, H~Okamoto, and Y~Tokura.
\newblock Ultrafast optical switching to a metallic state by photoinduced mott
  transition in a halogen-bridged nickel-chain compound.
\newblock {\em Physical review letters}, 91(5):057401, 2003.

\bibitem{rosch2008}
Achim Rosch, David Rasch, Benedikt Binz, and Matthias Vojta.
\newblock Metastable superfluidity of repulsive fermionic atoms in optical
  lattices.
\newblock {\em Physical review letters}, 101(26):265301, 2008.

\bibitem{sensarma2010}
Rajdeep Sensarma, David Pekker, Ehud Altman, Eugene Demler, Niels Strohmaier,
  Daniel Greif, Robert J{\"o}rdens, Leticia Tarruell, Henning Moritz, and
  Tilman Esslinger.
\newblock Lifetime of double occupancies in the fermi-hubbard model.
\newblock {\em Physical Review B—Condensed Matter and Materials Physics},
  82(22):224302, 2010.

\bibitem{eckstein2011}
Martin Eckstein and Philipp Werner.
\newblock Thermalization of a pump-excited mott insulator.
\newblock {\em Physical Review B—Condensed Matter and Materials Physics},
  84(3):035122, 2011.

\bibitem{ichikawa2011}
Hirohiko Ichikawa, Shunsuke Nozawa, Tokushi Sato, Ayana Tomita, Kouhei
  Ichiyanagi, Matthieu Chollet, Laurent Guerin, Nicky Dean, Andrea Cavalleri,
  Shin-ichi Adachi, et~al.
\newblock Transient photoinduced ‘hidden’phase in a manganite.
\newblock {\em Nature materials}, 10(2):101--105, 2011.

\bibitem{lenarvcivc2013}
Zala Lenar{\v{c}}i{\v{c}} and Peter Prelov{\v{s}}ek.
\newblock Ultrafast charge recombination in a photoexcited mott-hubbard
  insulator.
\newblock {\em Physical Review Letters}, 111(1):016401, 2013.

\bibitem{stojchevska2014}
L~Stojchevska, I~Vaskivskyi, T~Mertelj, P~Kusar, D~Svetin, S~Brazovskii, and
  D~Mihailovic.
\newblock Ultrafast switching to a stable hidden quantum state in an electronic
  crystal.
\newblock {\em Science}, 344(6180):177--180, 2014.

\bibitem{mitrano2014}
Matteo Mitrano, Giovanni Cotugno, SR~Clark, Rashmi Singla, Stefan Kaiser, Julia
  St{\"a}hler, R~Beyer, M~Dressel, Leonetta Baldassarre, Daniele Nicoletti,
  et~al.
\newblock Pressure-dependent relaxation in the photoexcited mott insulator
  et--f 2 tcnq: Influence of hopping and correlations on quasiparticle
  recombination rates.
\newblock {\em Physical review letters}, 112(11):117801, 2014.

\bibitem{werner2019}
Philipp Werner, Martin Eckstein, Markus M{\"u}ller, and Gil Refael.
\newblock Light-induced evaporative cooling of holes in the hubbard model.
\newblock {\em Nature communications}, 10(1):5556, 2019.

\bibitem{peronaci2020}
Francesco Peronaci, Olivier Parcollet, and Marco Schir{\'o}.
\newblock Enhancement of local pairing correlations in periodically driven mott
  insulators.
\newblock {\em Physical Review B}, 101(16):161101, 2020.

\bibitem{li2020}
Jiajun Li, Denis Golez, Philipp Werner, and Martin Eckstein.
\newblock $\eta$-paired superconducting hidden phase in photodoped mott
  insulators.
\newblock {\em Physical Review B}, 102(16):165136, 2020.

\bibitem{diehl2008}
Sebastian Diehl, A~Micheli, Adrian Kantian, B~Kraus, HP~B{\"u}chler, and Peter
  Zoller.
\newblock Quantum states and phases in driven open quantum systems with cold
  atoms.
\newblock {\em Nature Physics}, 4(11):878--883, 2008.

\bibitem{kraus2008}
Barbara Kraus, Hans~P B{\"u}chler, Sebastian Diehl, Adrian Kantian, Andrea
  Micheli, and Peter Zoller.
\newblock Preparation of entangled states by quantum markov processes.
\newblock {\em Physical Review A—Atomic, Molecular, and Optical Physics},
  78(4):042307, 2008.

\bibitem{coulthard2017}
JR~Coulthard, Stephen~R Clark, S~Al-Assam, Andrea Cavalleri, and D~Jaksch.
\newblock Enhancement of superexchange pairing in the periodically driven
  hubbard model.
\newblock {\em Physical Review B}, 96(8):085104, 2017.

\bibitem{zhang2020b}
XZ~Zhang and Z~Song.
\newblock Dynamical preparation of a steady off-diagonal long-range order state
  in the hubbard model with a local non-hermitian impurity.
\newblock {\em Physical Review B}, 102(17):174303, 2020.

\bibitem{zhang2021eta}
XZ~Zhang and Z~Song.
\newblock $\eta$-pairing ground states in the non-hermitian hubbard model.
\newblock {\em Physical Review B}, 103(23):235153, 2021.

\bibitem{yang2022dynamic}
XM~Yang and Z~Song.
\newblock Dynamic transition from insulating state to $\eta$-pairing state in a
  composite non-hermitian system.
\newblock {\em Physical Review B}, 105(19):195132, 2022.

\bibitem{zhang2021topologically}
KL~Zhang and Z~Song.
\newblock Topologically protected two-fluid edge states.
\newblock {\em Physical Review B}, 104(18):184515, 2021.

\bibitem{wang2024flat}
R~Wang and Z~Song.
\newblock Flat band and $\eta$-pairing states in a one-dimensional moir{\'e}
  hubbard model.
\newblock {\em Chinese Physics Letters}, 41(4):047101, 2024.

\bibitem{zhang2015topological}
G~Zhang and Z~Song.
\newblock Topological characterization of extended quantum ising models.
\newblock {\em Physical review letters}, 115(17):177204, 2015.

\bibitem{zhang2017majorana}
G~Zhang, C~Li, and Z~Song.
\newblock Majorana charges, winding numbers and chern numbers in quantum ising
  models.
\newblock {\em Scientific Reports}, 7(1):8176, 2017.

\bibitem{vafek2017entanglement}
Oskar Vafek, Nicolas Regnault, and Bogdan~Andrei Bernevig.
\newblock Entanglement of exact excited eigenstates of the hubbard model in
  arbitrary dimension.
\newblock {\em SciPost Physics}, 3(6):043, 2017.

\bibitem{spielman2024quantum}
Sarah~E Spielman, Alicia Handian, Nina~P Inman, Thomas~J Carroll, and Michael~W
  Noel.
\newblock Quantum many-body scars in few-body dipole-dipole interactions.
\newblock {\em Physical Review Research}, 6(4):043086, 2024.

\bibitem{blanes1998magnus}
Sergio Blanes, Fernando Casas, JA~Oteo, and J~Ros.
\newblock Magnus and fer expansions for matrix differential equations: the
  convergence problem.
\newblock {\em Journal of Physics A: Mathematical and General}, 31(1):259,
  1998.

\end{thebibliography}
\end{document}